\documentclass[preprint, 12pt,
 amsmath,amssymb,
 aps, 
 prl
]{revtex4-2}

\usepackage{graphicx}
\usepackage{dcolumn}
\usepackage{bm}
\usepackage{soul}
\usepackage[utf8]{inputenc}
\usepackage[T1]{fontenc}
\usepackage{mathptmx}
\usepackage{etoolbox}
\usepackage{hyperref}

\usepackage{color}

\newcommand{\RR}{{\mathbb{R}}}

\newcommand{\QQ}{{\mathbb{Q}}}
\newcommand{\PP}{{\mathbb{P}}}
\usepackage{xcolor}

\makeatletter
\def\@email#1#2{%
 \endgroup
 \patchcmd{\titleblock@produce}
  {\frontmatter@RRAPformat}
  {\frontmatter@RRAPformat{\produce@RRAP{*#1\href{mailto:#2}{#2}}}\frontmatter@RRAPformat}
  {}{}
}%
\makeatother
\begin{document}


\title[Measuring Dimension]{{Using Wavelet Decomposition to Determine the Dimension of Structures from Projected Images}}
\author{Svitlana Mayboroda}
\affiliation{ETH Zurich, Department of Mathematics, Rämistrasse 101 8092, Zürich, Switzerland}

\author{David N Spergel}%
 \email{dspergel@flatironinstitute.org}
\affiliation{ 
Flatiron Institute, 162 Fifth Avenue NY NY 10011 USA
}%

\date{\today}

\begin{abstract}
{Mesoscale structures can often be described as fractional dimensional across a wide range of scales. We consider a $\gamma$-dimensional measure embedded in an N dimensional space and discuss how to determine its dimension, both in N-dimensions and projected into D-dimensions. It is a highly non-trivial problem to decode the original geometry from lower dimensional projection of a high-dimensional measure. The projections are space-filling, the popular box-counting techniques do not apply, and the Fourier methods are contaminated by aliasing effects. In the present paper we demonstrate that we are not observing objects from a special direction, projection in a wavelet basis is remarkably simple: the wavelet power spectrum of a projected $\gamma$ dimensional measure is $P_j= 2^{-j \gamma}$. This holds regardless of the embedded dimension, $N$, and the projected dimension, $D$. This approach could have potentially broad applications in data sciences where a typically sparse matrix encodes lower dimensional information embedded in an extremely high dimensional field and often measured in projection to a low dimensional space. Here, we apply this method to JWST and Chandra observations of Cas A We find that the emissions can be represented by projections of mesoscale substructures with fractal dimensions varying from $\gamma= 1.7$ for the warm CO layer observed by JWST, up to $\gamma= 2.5$ for the hot X-ray emitting gas layer in the supernova remnant. The resulting power law indicates that the emission is coming from a fractal dimensional mesoscale structure in expanding supernova remnant.}
\end{abstract}

\maketitle
\par \vfill \eject

\section{1. Introduction}
\label{sec:level1}

Observations of nebulae, molecular clouds, and other astronomical objects reveal intricate structures.  In this paper, we discuss how to infer the fractal dimension of a structure in three-dimensional space from projected images.  While this work has been motivated by examples from astronomy, our methods are applicable to a broad range of fields.

Many objects show scaling behavior over many orders of magnitude, \textcolor{black}{often generated by hydrodynamical or magneto-hydrodynamical turbulence \cite{Kritsuk2007,Schekochihin2022}.  Astronomical observations show evidence for turbulence spanning up to 12 orders of magnitude\cite{Chepurnov2010,yuen2022}.} If we could accurately measure the dimension of these structures in 3D,  then we could obtain new insights into the underlying physics.  The dimension of a level set of a passive scalar in a turbulent flow  appears to depend on the Damk\"ohler number \cite{Chatakonda}: flames develop a 7/3 fractal if the chemical reaction time (or equivalently the cooling time) is short compared to the Eddy turnover time of the smallest eddies  and 8/3 if the Eddy turnover time scale is comparable to the chemical reaction time.   Simulations \cite{2020ApJ...894L..24F} suggest that the Kelvin-Helmholtz instability generates a thermal emission layer  with dimension 5/2 while simulations  of molecular clouds \cite{2019MNRAS.487.2070B} show an index that varies with Alfvenic Mach number. If we could infer the dimension of a structure in three dimensions, we would have important insights into the physics that determines the observed mesoscale structures.

Astronomers observe the projected images of 3-dimensional object.   If the emission at a given frequency, $\nu$,\textcolor{black}{ is optical thin}, then the  surface brightness of the image is projection of the emissivity of the object, $\epsilon_\nu$:
\begin{equation}
    \Sigma_\nu(x,y) = \int dz \ \epsilon_\nu(x,y,z).
\label{eq:sigma} 
\end{equation}

Our underlying physical model is that we are observing emission from a structure that can be approximated  as a $\gamma$-dimensional surface on its mesoscale, $r_{min} < r < r_{max}$.  Here, $r_{min}$ is the thickness of the structure and $r_{max}$ is the outer scale of the instability that creates the mesoscale phenomenon.  Note that equation (1) implies that we are interested in measures and their projections, not sets. Indeed, when $\gamma>2$, the projection of a set would be typically space-filling, that is, two-dimensional.  Observations of a projected set provide no information about the initial structure.    Thus, it is important that we are observing a projection (e.g., equation \ref{eq:sigma}). In this paper, we will treat not only the projection from 3 to 2 dimensions but a more general case of the projection from $N$ to $D$ dimensions.

What is the most effective way of describing the fractal dimension of a measure on the mesoscale? Even with the access to the full structure, not just its projection, this is a very intricate question, and the traditional toolbox is surprisingly misleading. The fundamental problem is that numerical methods developed for ``classical" fractals do not accurately capture the mesoscale phenomena in highly disconnected sets under consideration.
  While  box counting methods or equivalently, perimeter-area scalings \cite{Bazell1988}, are widely used to describe sets in fields ranging from engineering \cite{Wu2020} to plant science \cite{Bouda2016} to ophthalmology \cite{LiebovitchToth1989} to astronomy \cite{2019MNRAS.487.2070B,Marchuk2021} we argue that this approach fails even for the famous problem of determining the properties of a coastline. 
This method counts the number of boxes, $K$, in $\RR^N$ that contain $\mu$ as a function of scale, $L$ and then defines the Minkowski, or box-counting, dimension as $d\log K/d\log L$.  This approach is problematic \cite{Falconer1990}, particularly if $\mu$ is not a continuous surface. An important caveat is that in many applications $\mu$ is fragmented into bubbles.  While it may be tempting to introduce a ``filling factor" to count the boxes containing $\mu$, this is an ill-defined concept: in the limit,  the points in $\mu$ are a set of measure zero in $\RR^N$ and the filling factor, the volume of the box filled by $\mu$, is zero.  One could say that a bubble is not a point, but when viewed on the mesoscale, a bubble or tidal pool appears as "dust" and box counting breaks down in the presence of "dust".  As an illustration of its failure, consider the rational numbers  $\QQ \subset \RR$: box counting yields a dimension 1 for $\QQ$
and dimension $N$ for $\QQ^N$, that is, it does not distinguish between the set of rational numbers on a line and a line itself.  The Hausdorff dimension of the set of real numbers is zero.
If you are a marine biologist who wants to determine the surface area of the tidal zone, you do not want to exclude tidal pools.   Similarly, isolated bubbles are not properly counted in analyses of molecular clouds or measurements of the area of a flame front in studies of combustion. 
 In \S 2, we show that wavelets are a sparse and effective description of a measure that provides a more robust approach than box counting on sets.

An even bigger problem, central to the present paper, is that in astronomy one often only has access to projected images. If one is projecting to $\RR^D$ a set of dimension larger than 2, the result is typically two-dimensional, space-filling, and box counting fails for much more trivial reasons than above: it blindly yields projected dimension 2. The Fourier methods, if attempted, would be contaminated by aliasing. 
Moreover, we only observe one projection at a time, as opposed to the view from all angles. Thus, the methods of reconstructing the data from projections at all angles used, e.g., in CT, do not apply.  This problem is the focus of \S 3, where the main contributions of the paper are described.  We develop a new method to infer fractal dimension of a mesoscale structure in $\RR^N$ from observations of projected images in $\RR^D$, under appropriate physical assumptions.

In \S 4, we apply this method to an image of the emission from warm CO derived from  JWST image of Cas A at F356W and F444W \cite{Milisavljevic_2024,Rho2024} and to X-ray images of Cas A\cite{Patnaude2009}.  
\section{2. Measuring Fractal Dimension}
\label{sec:fractal}

This section focuses on methods for describing the dimension of a measure with either wavelets or Fourier methods.  We refer the reader to classical texts, e.g.,  \cite{Falconer1990}, for rigorous definitions and analysis of some of the mathematical concepts used below.

\textcolor{black}{As elucidated above, Minkowski (box-counting) dimension is not the most effective description of the geometry of disconnected dimensions so we focus on Hausdorff dimension as a tool to describe these structures.}
We consider a set in ${\mathbb R}^N$ with a fractional dimension. We will refer to it as a fractal, although we do not assume self-similarity, simply that it exhibits the same dimension across a range of scales.  The Hausdorff dimension, $\gamma$, of  a set, is the supremum of $s$ so that  the Riesz potential,
\begin{equation}
    I_s(\mu) = \int \frac{d\mu_x d\mu_y}{|x-y|^s},
\end{equation}
is finite for some non-trivial and finite Borel measure $\mu$ on the set. 

Wavelets provide an effective language for describing the dimension of a measure.  We pick a wavelet basis and
let $\phi_{jm}=2^{jN/2}\phi((x-x_{jm})/2^{-j})$ denote the wavelets concentrated on a cube centered at $x_{jm}\in 2^{-j}{\mathbb Z}^N$ of 
the side length $2^{-j}$. We make sure that wavelets are $L^2$ normalized, that is, $\int_{\RR^N} |\phi_{jm}|^2=1$, and denote the amplitude of a wavelet coefficient at scale $2^{-j}$ centered at position $x_{jm}$ by
\begin{equation}
    a_{jm} = \int \phi_{jm} \,d\mu.
\end{equation}

Because fractal dimensional surfaces are sparse and typically not space-filling, wavelets are a sparse representation. 
In this language, most of the wavelet coefficients (concentrated on boxes that do not intersect the support of $\mu$) are close to zero. When, on the other hand, the box corresponding to $\phi_{jm}$ has an ample intersection with ${\rm supp}\, \mu$, generally the amplitude of the corresponding coefficient $a_{jm}$ scales as $2^{-j\gamma}2^{jN/2} $ (see, e.g., \cite{Hazewinkel}, for an analogous computation). Given that $\mu$ is $\gamma$-dimensional in ${\mathbb R}^N$, the proportion of boxes with a non-trivial contribution at every scale is $2^{j(\gamma-N)}.$ Working at mesoscales $r_{min}<r<r_{max}$ corresponds to considering wavelets with $-j_{max}<j<-j_{min}$.  On these scales, the wavelet power spectrum has a power-law behavior:
\begin{equation}
    P_j = \frac{1}{2^{jN}} \sum_m |a_{jm}|^2 \propto 2^{j(\gamma-N)} 2^{j(N-2\gamma)} \propto 2^{-j\gamma}.
\end{equation}

Alternatively, we can describe the measure in a Fourier basis:
\begin{equation}
    \widehat \mu(k) = \int  \exp(ikx) d\mu(x), \quad k\in {\mathbb R}^N.
\end{equation}
Now, the Riesz potential becomes:
$$I_s(\mu)=\int |k|^{s-N} |\widehat \mu(k)|^2\, dk. $$
Hence, bringing us back to the power spectrum, if we assume the power behavior of $|\widehat \mu(k)|^2$,  the dimension can be viewed as $\inf\{\tau:\,|\widehat \mu(k)|^2 \leq C|k|^{-\tau}$\}, or more informally, $\gamma$ is the mesoscopic dimension if $|\widehat \mu(k)|^2 \approx |k|^{-\gamma}$ for $\frac{1}{r_{max}}<|k|<\frac{1}{r_{min}}$. The description of the measure in Fourier modes is, however, a less sparse representation as the information is spread through Fourier space.


\section{\label{sec:project} 3. Measuring Fractal Dimension from Projected Images}

How do we deduce the dimension of the original set, $\gamma$, looking at its projection on ${\mathbb R}^D$? When $\gamma<D$, it almost surely projects into a set of dimension $\gamma$ again, according to Marstand's 1950 theorem. When $\gamma>D$, which is a typical situation discussed in the present paper, the situation is much more delicate. The projection of a set itself would typically be simply $D$-dimensional, space-filling, giving us no information. However, we have richer data. We actually project a {\it measure} not a set, in the sense that we retain the information of the number of intersections with the set when projecting, but then the challenge is how to take advantage of this information.

Part of the problem is that projection always loses information. Physical systems can have preferred directions determined by either large-scale gravitational fields (e.g., the Earth for oceanographers or atmospheric physicists) or large-scale magnetic fields. If the properties of the set depend on direction, a projection parallel or perpendicular to the symmetry direction yields very different projected images. Luckily,   astronomical observations are typically not along a symmetry axis of the system.  \textcolor{black}{We can
think of this as a form of the Copernican principle that we do not observe the objects in the universe
from a special position or orientation.}

Projecting the ${\mathbb R}^N$ wavelet decomposition $\sum_{jm} a_{jm} \phi_{jm}$ to ${\mathbb R}^D$, we use the property that ``horizontally oriented" 3D wavelets $\phi_{jm}$, $m\in 2^{-j}{\mathbb Z}^N$, project identically onto 2D wavelets modulo a renormalization:  $\PP \phi_{jm}=2^{-j\,\frac{N-D}{2}}\phi^\sharp_{jm^\sharp}$, $m^\sharp\in 2^{-j}{\mathbb Z}^D$. 
The ``vertically oriented" wavelets project to zero. This would be a problem if horizontal or vertical were a preferred direction, but since we assume that there is none, the information contained in all coefficients $a_{jm}$ is the same. 
Hence, we obtain a decomposition of the projected measure, 
\begin{equation}\label{phiproj}
\sum_{jm^\sharp} a_{jm^\sharp}\phi_{jm^\sharp}, 
\end{equation} 
where $a_{jm^\sharp}$ is the sum of the coefficients $a_{jm}$ for $m$ above $m^\sharp$ times $2^{-j\,\frac{N-D}{2}}$.  For the astronomical images described in the introduction,
equation (\ref{phiproj}) is essentially an expansion of $\Sigma(x,y)$.

\textcolor{black}
{In $N$ dimensional space on scale $j$, there are $2^{jN}$ wavelets in a box of size 1. For a $\gamma$ dimensional set, a fraction of 
roughly $2^{-j(N-\gamma)}$ of the $a_{jm}$'s are non-zero.  These $2^{j\gamma}$ non-zero wavelets project into 
$2^{jD}$ wavelets in the projected image.  Thus,  each $a_{jm^\sharp}$ is a sum of typical $M$ amplitudes 
\begin{equation}
    M(j) = 2^{j(\gamma-D)}  = \left(\frac{1}{r}\right)^{(\gamma-D)},
\end{equation}
where $r$ is the wavelet scale associated with the $j$-th wavelets. 
If the coefficients of the modes sum incoherently \footnote{We thank Guy David for suggesting this $M^{1/2}$ scaling}, then
\begin{eqnarray}
    a_{jm^\sharp} &\propto& M^{1/2} 2^{-j\gamma} 2^{jN/2}2^{-j\,(N-D)/2}  \propto 2^{j(\gamma-D)/2} 2^{-j\gamma}2^{j\,D/{2}}  \nonumber \\ &\propto& 2^{-j\gamma/2}. \label{aproj}
\end{eqnarray}
Thus, the dimension of the initial image can be deduced from its $D$-dimensional wavelet decomposition (equation (\ref{phiproj})) by using the scaling law (equation (\ref{aproj})). }

\begin{figure*}[h]
  \centering
  {\includegraphics[width=.45\textwidth]{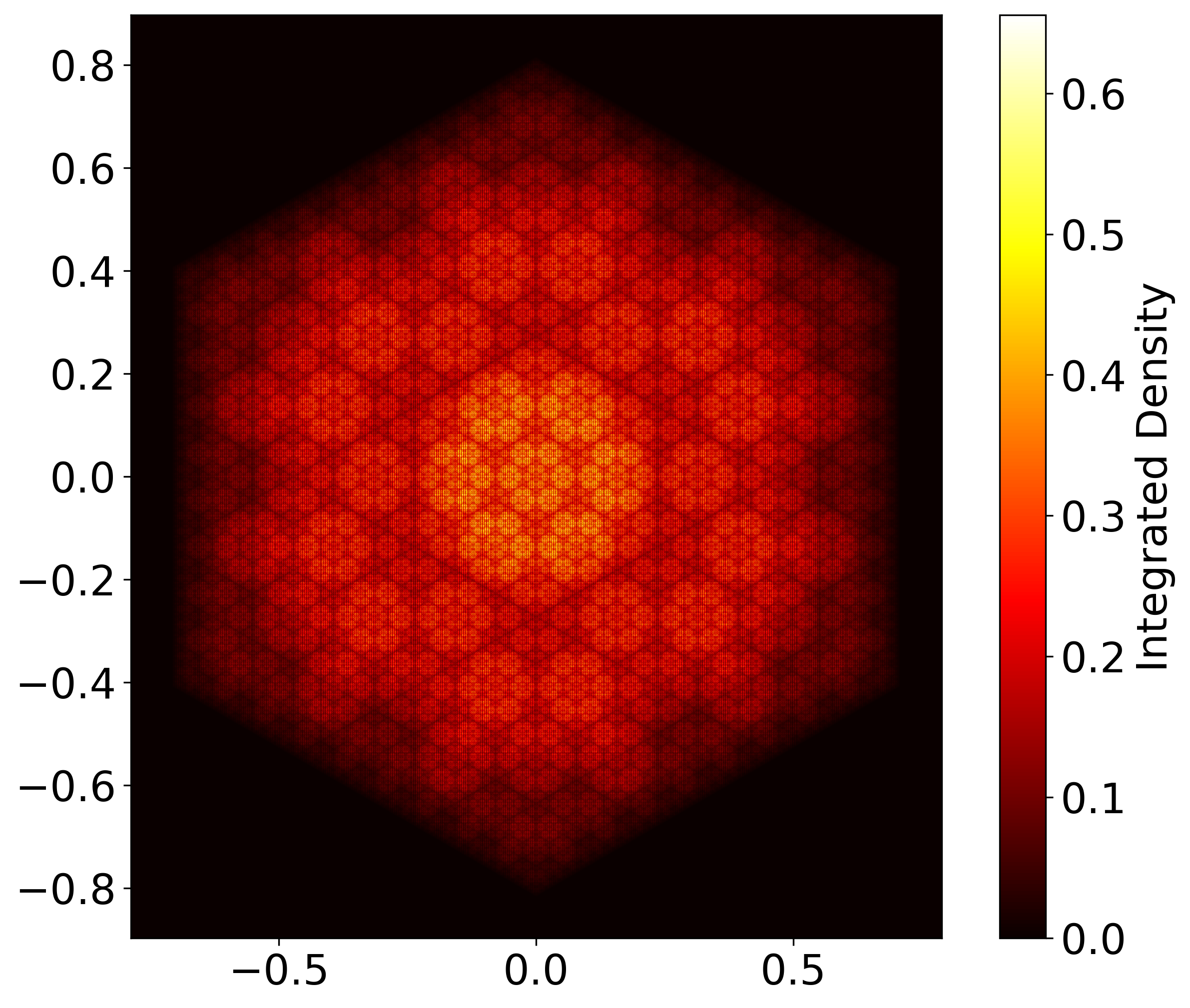}}\quad
  {\includegraphics[width=.45\textwidth]{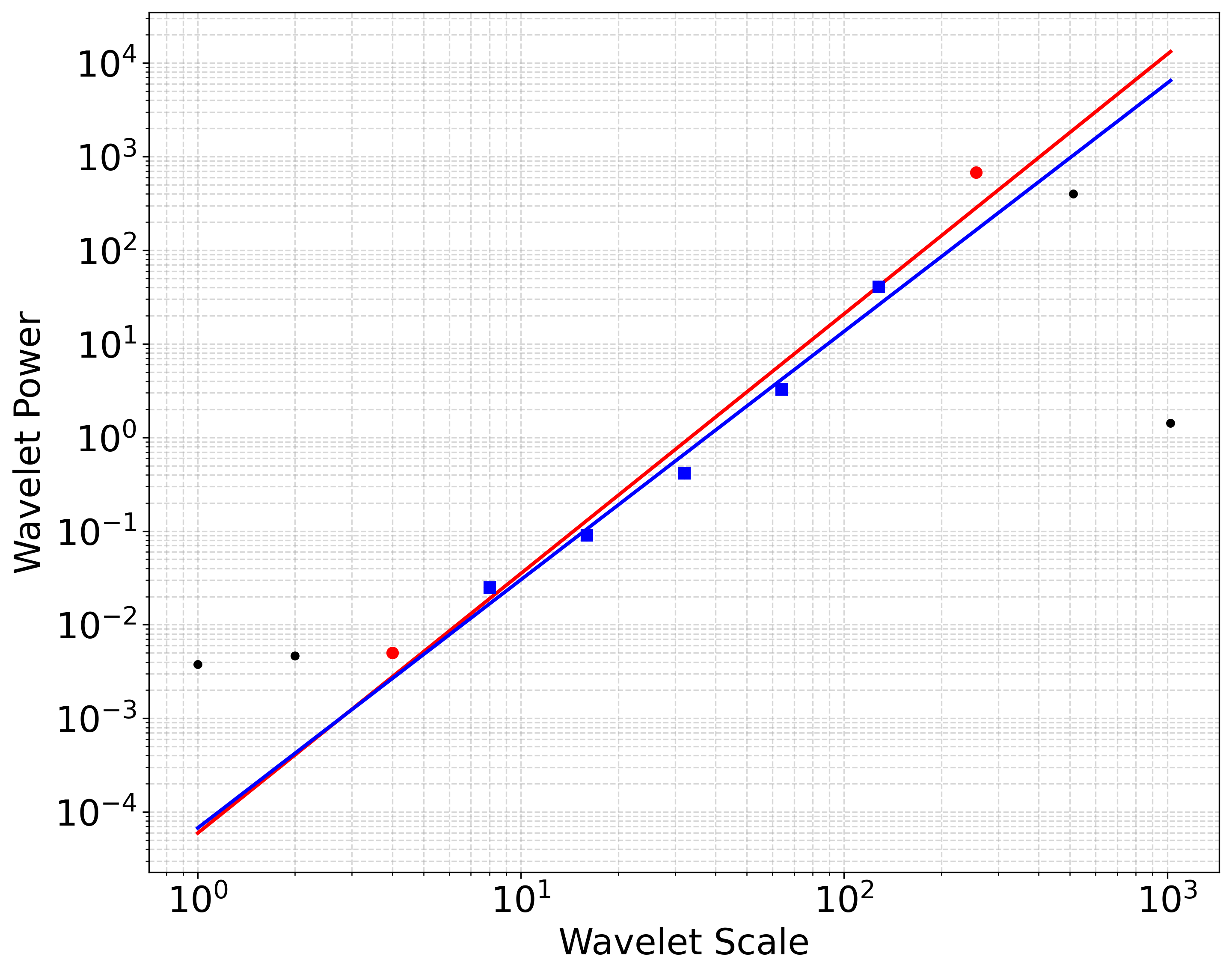}}\\
  {\includegraphics[width=.45\textwidth]{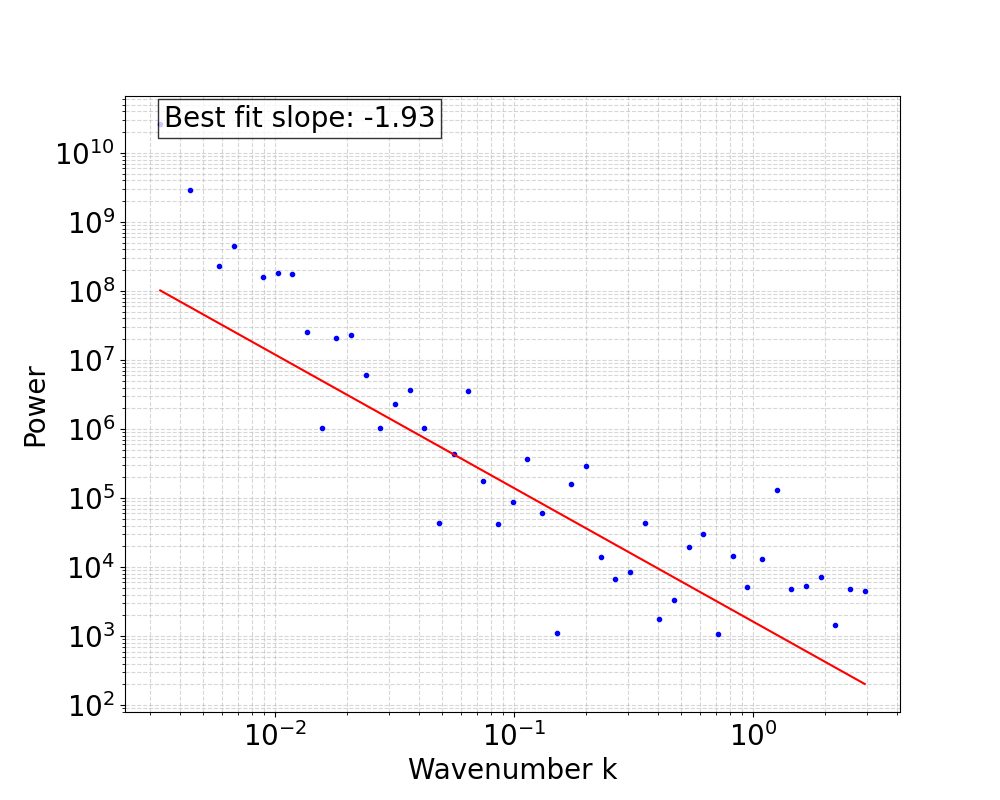}}\quad
  {\includegraphics[width=.45\textwidth]{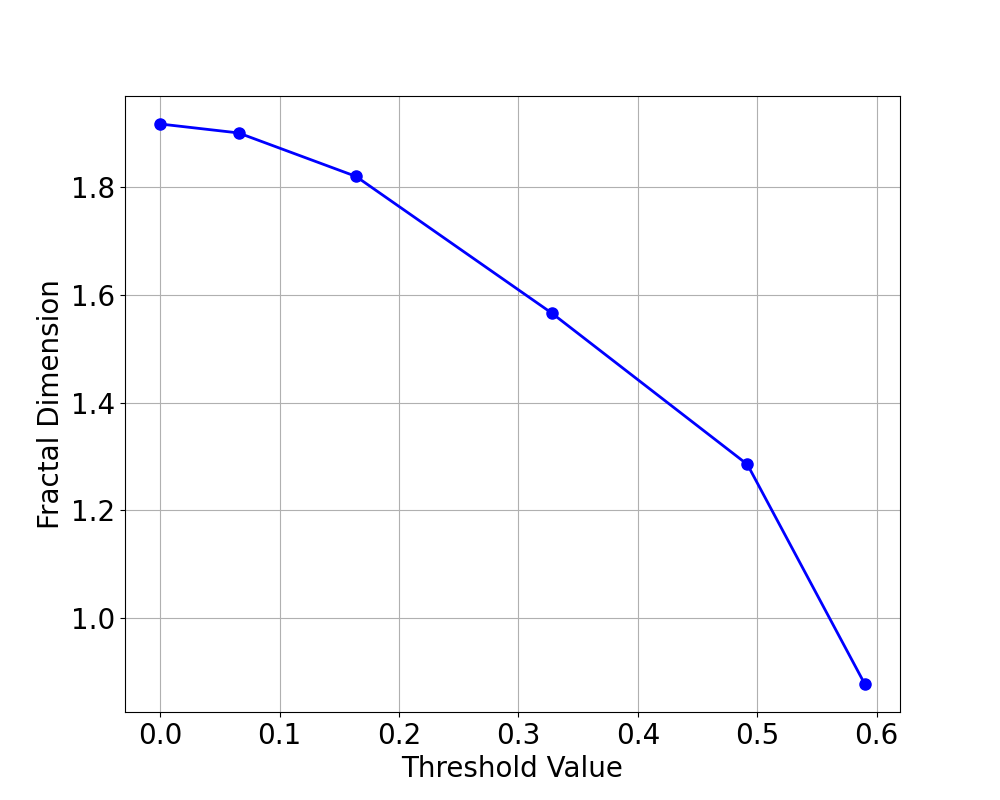}}

    \caption{\textcolor{black}{
    The upper left panel shows the projected density of a six level Menger sponge, a fractal of dimension $\gamma \sim 2.767$.  The projection is along a line that is  30 degrees off of the principal axis.  The upper right panel shows the wavelet amplitude versus scale and fits for two different ranges: a fit from 4-256 yields a slope of 2.773 and a fit of 8-128 yields a slope of 2.652.
    The lower left panel show the power spectrum of the Fourier transform of the image.    Because of the finite size of the object, the FFT method yields 1.912 for the fractal dimension. When tested on Note that  a Menger sponge embedded in a periodic box ($T^3$), it yields a value close to the correct value. The lower right panel shows  the inferred dimension using a threshold method as a function of threshold.   The value in  the  figure should be compared to $\gamma -1 = 1.767$.
        \label{fig:menger}}}
\end{figure*}

\begin{figure*}[h]
    \centering{\includegraphics[width=0.9\textwidth]{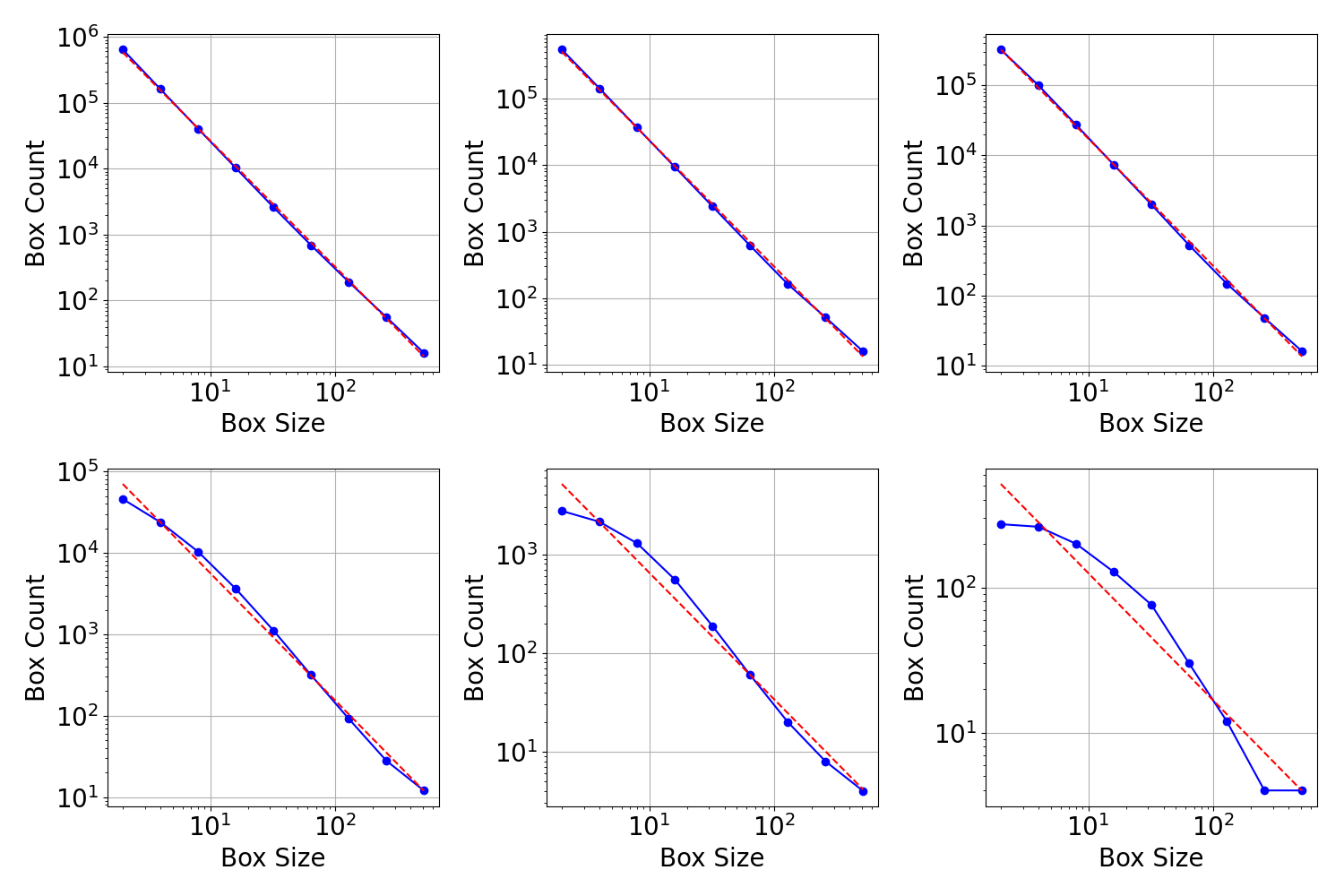}
    \caption{{\textcolor{black} {This figure shows the box count versus box size for the projected Menger sponge.  For box thresholds set at 0.0, 0.1, 0.2, 0.5, 0.8 and 0.9, the best fit lines (shown as dashed red) have slope 1.902, 1.894, 1.803, 1.521, 1.175 and 0.870.  The correct answer should be $\sim 1.727$: while the box counting method can yield convincing power laws, it produces incorrect results when applied to projected density fields.
    \label{fig:box_count}}}}}
\end{figure*}

\textcolor{black}{
Figure (\ref{fig:menger}) shows a projection of a Menger sponge at an angle of 30$\^o$.  The Menger fractal has dimension $\log_3(20) \simeq 2.727$. The figure shows the success of the wavelet method and the limitations of Fourier and box counting methods: both yield incorrect results.  Figure (\ref{fig:box_count} shows the inferred fractal dimension versus threshold: box counting can yield a convincing power law for some threshold values but an incorrect result: this may have led to erroneous conclusions in the literature.}

Many astronomical objects \textcolor{black}{are projection of complex shapes.}  At each wavelet position, the path length, $R_{m^\sharp}$, through the three dimensional structure varies.  These leads
to a variation in 
\begin{equation}\label{shell1}
    M(j,m^\sharp) = \left(\frac{R_{m^\sharp}}{r}\right)^{\gamma-D},
\end{equation}
However, when we average across the image, this variation does not change the power law scaling. 

\textcolor{black}{
Averaging over the image, we can define a characteristic thickness,
$R_{\rm{eff}}$:
\begin{equation}\label{eqshell}
    \overline{M}_{projected}(j) =\frac{1}{2^{jD}} \sum_{m^\sharp}\left(\frac{R_{m^\sharp}}{r}\right)^{\gamma-D} \propto \left(\frac{R_{\rm{eff}}}{r}\right)^{(\gamma-D)} \propto 2^{j(\gamma-D)}.
\end{equation}
Note that $R_{\rm{eff}}$ can be defined independently of $j$ (only depending on $\gamma$, $D$, and the shape of the shell). Then again the computation yields the scaling of the wavelet power spectrum: }
\begin{equation}\label{shell2}
  P_{j}^{wavelet} = \frac{1}{2^{jD}}\sum_{m^\sharp} a_{jm^\sharp}^2 \propto 2^{-j\gamma}.
\end{equation}
\textcolor{black}{
As long as the path length is large compared to the minimum scale size, the length, $R_{m^\sharp}$ ultimately plays no role in \eqref{shell1}--\eqref{shell2}. The wavelets mod out the global structure effortlessly and the dominating shape is not important. This makes it clear why the proposed method is superior to  the Fourier decomposition.}

Indeed, for a homogeneous uniform slab in  \eqref{phiproj}--\eqref{aproj}, the analysis could be done in Fourier space.
Projection in physical space is a slice (restriction) in Fourier space.  If we apply the ``Copernican principle" and assume that we are not seeing an object from a preferred directions, then the power spectrum of the image in $D$ dimensions is just a slice through the $N$ dimensional power spectrum. For a projected measure,
\begin{equation}
|\widehat \mu^\sharp (k_D)|^2 = C k_D^{-\gamma} \label{power_proj}
\end{equation}
where $k_D$ is a vector in $D$ dimensional space whenever for the original measure $|\widehat \mu(k_N)|^2 = C k_N^{-\gamma}$. 

However, an analog of  \eqref{phiproj}--\eqref{aproj} would suffer from the dominating aliasing effects (see \S 4): one would need to know what is the shape of the initial shell and painfully extract delicate dimension information from an analogue of \eqref{shell2}, much less explicit in the Fourier representation.
Thus while projection is conceptually simplest in Fourier space, Fourier analysis is poorly suited as even isotropic processes can be inhomogeneous in projection due to these projection effects.  As seen in the figure, the Fourier analysis yields the incorrect slope without aliasing corrections.


\section{4. Application:  Cas A\label{sec:CasA}}

\begin{figure*}[h]
  \centering
  {\includegraphics[width=0.9\textwidth]{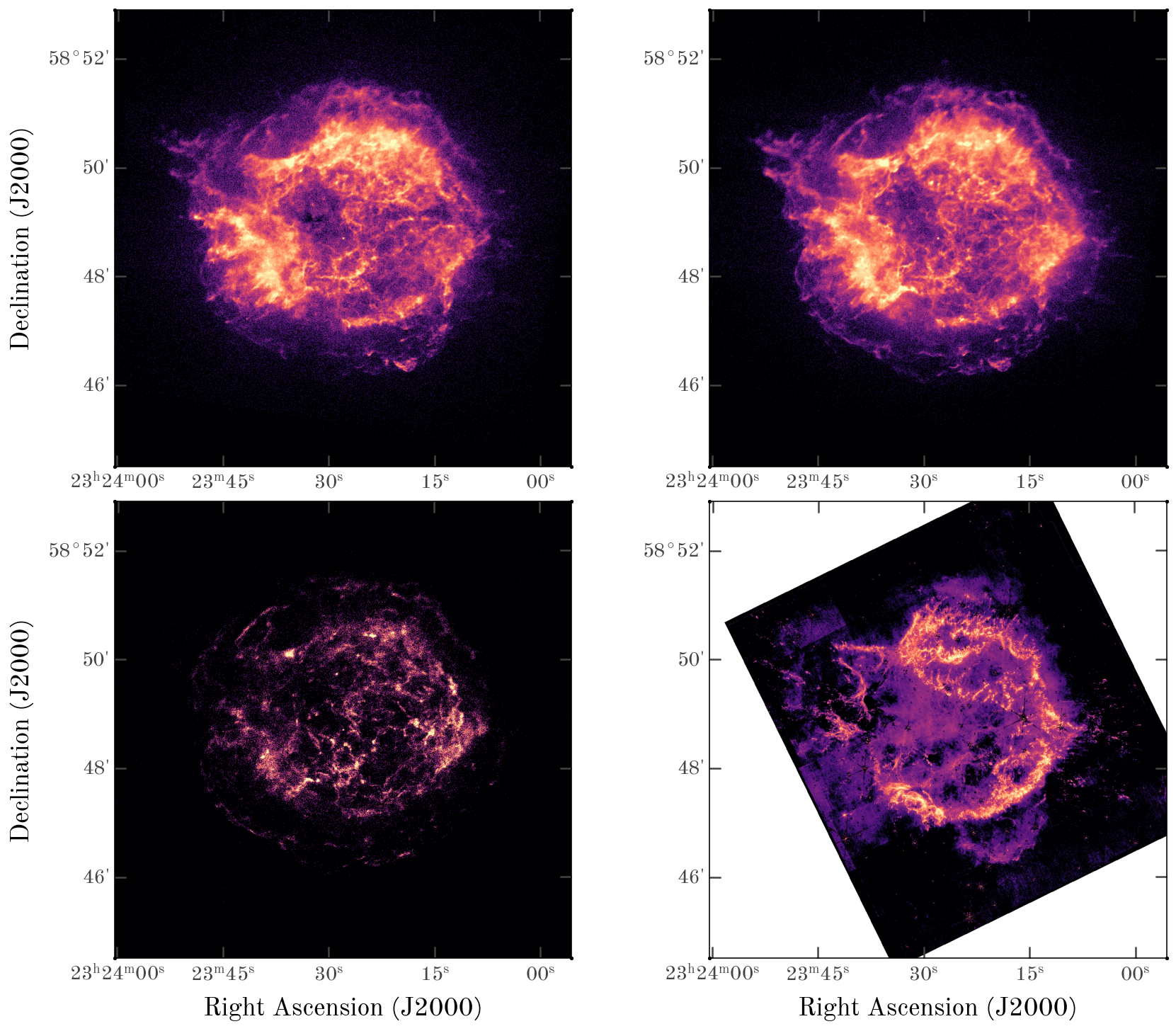}
  \caption{\textcolor{black}{X-ray and Infrared images of Cas A.   The first three images are Chandra X-ray images at 0.5-1.5 keV, 1.5-3.0 keV, and 4.0-6.0 keV.  The lower right figure is  derived \cite{Rho2024} by subtracting the F365W JWST NIRCAM image from F444W JWST NIRCAM image to remove the synchrotron emission.  We display the image of the data provided by Tea Temim from the JWST Cas A survey team.}}
  \label{fig:CasA}}
\end{figure*}

\begin{figure*}[h]
  \centering

  {\includegraphics[width=.45\textwidth]{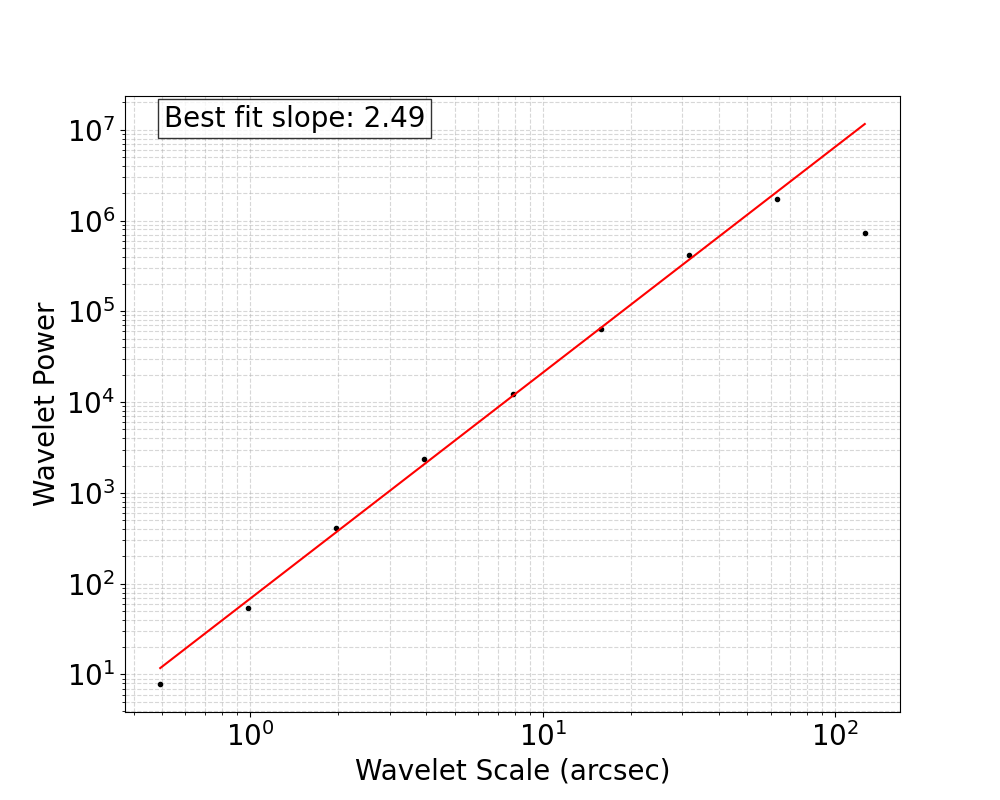}}\quad
  {\includegraphics[width=.45\textwidth]{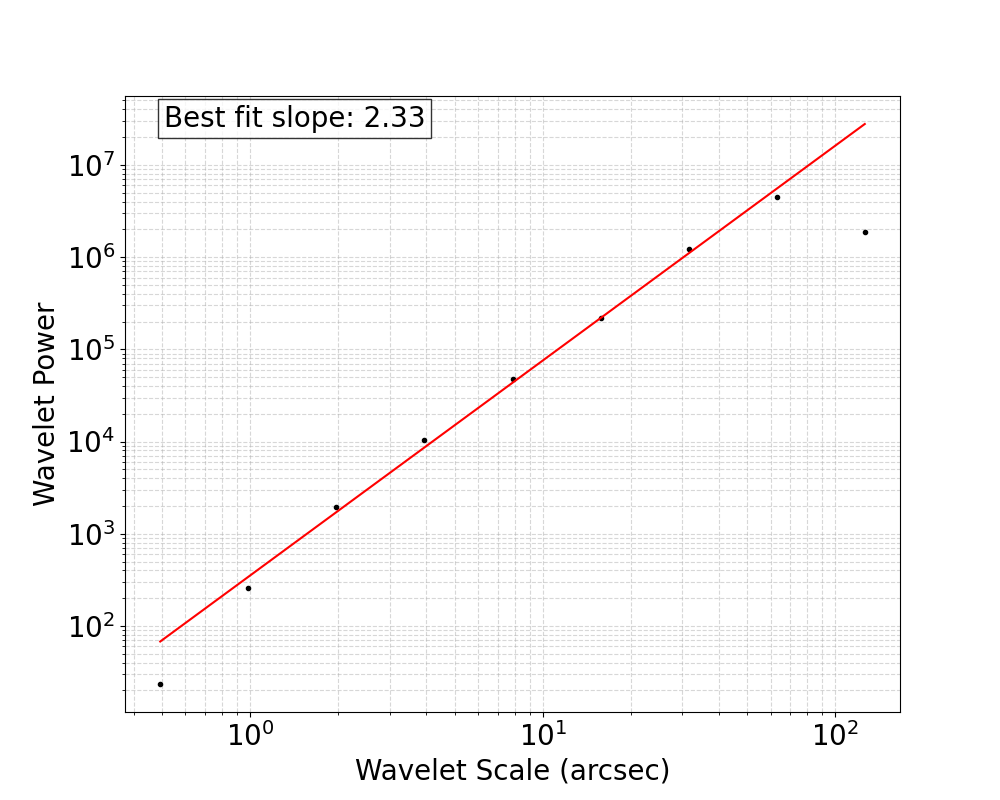}}\\
  {\includegraphics[width=.45\textwidth]{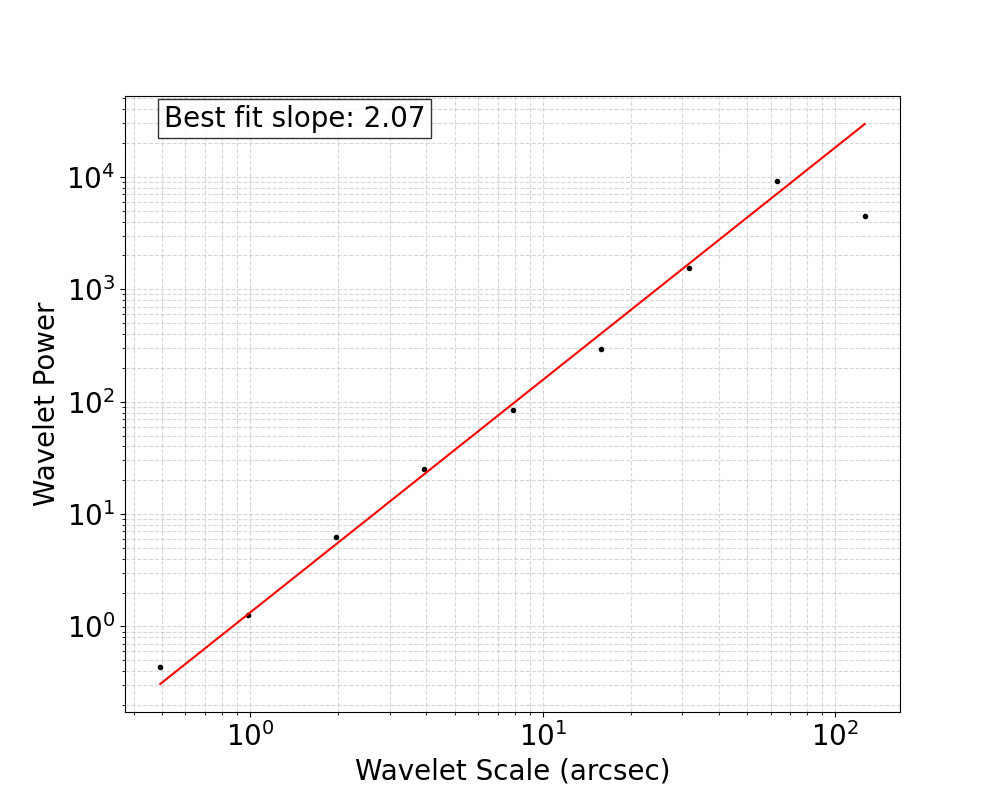}}\quad
  {\includegraphics[width=.45\textwidth]{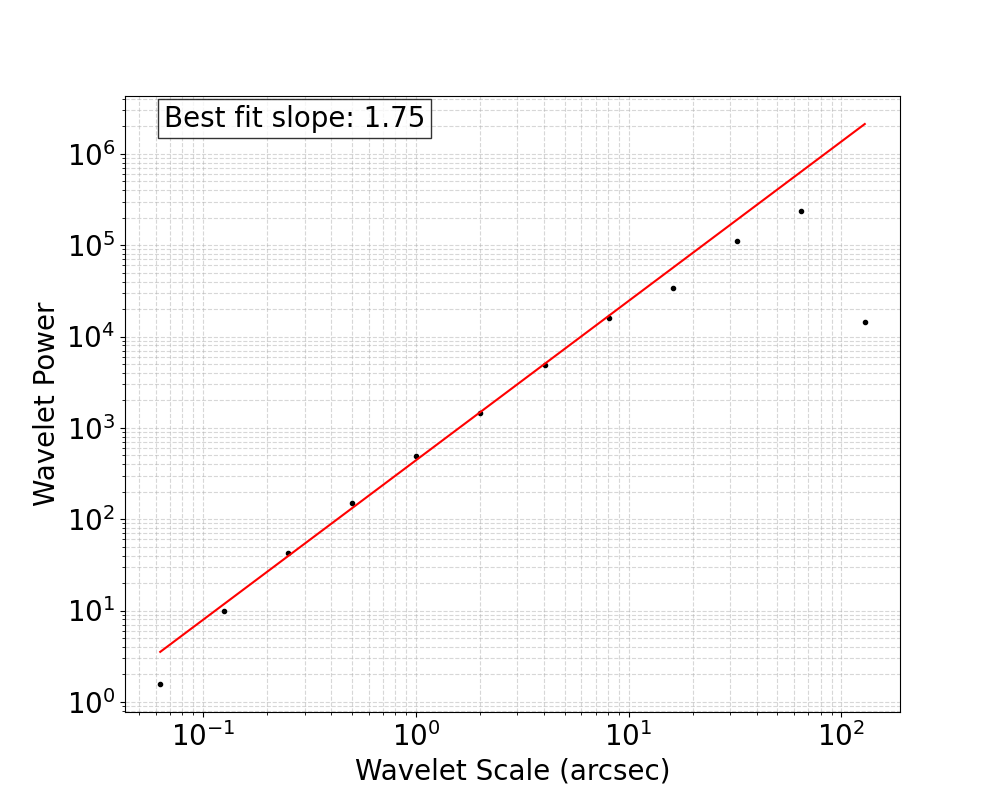}}
    \caption{Wavelet Power Spectrum of 2D Image of Cas A for gas at different frequencies.  Here, we use Ricker wavelets in the analysis. 
    The upper left is for 0.5-1.5 keV X-ray emission.  The upper right  is for 1.5-3.0 keV X-ray emission and the lower left is for 4.0-6.0 keV.  The lower right is for the cleaned JWST image. The break from the power law at the largest scales is 
    due to the fractal description of the emitting surface failing at scales comparable to the radius of Cas A.  Because of JWST's better angular resolution, the lower right spectrum has a larger dynamic range.
        \label{fig:Wavelet}}
\end{figure*}
In this section, we apply these methods to a mosaiced NIRCAM JWST image of Cas A  \cite{Milisavljevic_2024} and to Chandra image at 0.5-1.5, 1.5-3.0, and 4.0-6.0 keV \cite{Patnaude2009}.  This 
nearby and young ($\sim$ 350 yr) core-collapse supernova remnant is a particularly interesting case for this study.  There is high quality X-ray and near-IR observations that probe the forward and reverse shocks.  For over 60 years, physicists have suggested that these shocks are subject to Rayleigh-Taylor instabilities \cite{Gull1973}.  {\textcolor{black} While wavelets have been used   to characterize the substructure in X-ray remnants \cite{Lopez2011}, they have not been used to measure fractal dimension.}
Emission at F444W is dominated by warm CO and synchrotron emission, while emission at F365W is dominated by synchrotron \cite{Rho2024}. In this section, we analyze an image shown  in Figure (\ref{fig:CasA}) that is  derived by subtracting the F365W observations from F444W to remove the synchrotron emission \cite{Rho2024}. {\textcolor{black} We mask negative flux pixels in the synchrotron-subtracted maps provided by Tea Temim from the \cite{Milisavljevic_2024} analysis.} These {\textcolor{black} masked} pixels are either at the positions of stars or outside the reverse shock. The CO emission in the image is tracing the distribution of warm gas ($\sim 1000$K) in a reverse shock in the remnant \cite{Biscaro2014}.
This reverse shock is unstable to Rayleigh-Taylor instabilities and is expected to develop
a multi-scale fractal-like structure \cite{Blondin2001}.

        \label{fig:Wavelet_small}
Figure (\ref{fig:Wavelet}) shows the wavelet power spectrum, $P_j$, as a function of scale.  The wavelet transform of the image shows remarkable power law behavior across a wide range of scales with a spectral index depending on the wavelength.  This power law scaling suggests that the description of the image as the projection of fractal emitting layer is consistent with the data.  For the CO gas, the
slope of the wavelet power spectrum is 1.7.  For the X-ray emitting gas, the slope of the power spectrum
peaks near 2.5, consistent with simulations of instabilities of a thermal radiative mixing layer \cite{2020ApJ...894L..24F}.  It will be interesting to compare
 numerical studies of the development of instabilities in young supernova remnants \cite{Mandal2023} to these measurements from the observations.\textcolor{black} { Since fractal dimension measurements are a non-linear statitics, uncertainty analysis and model comparison  will require a full forward model that starts with realization of the theoretical model and incorporates measurement effects including resolution, foreground uncertainty removal and noise.  We hope that this paper motivates the use of fractal dimension as a tool for comparing theory and observation.}
 
\section{Conclusions}
In this paper, we establish a new method for analyzing images and determining the dimension of projected mesoscale measure, set up in the framework of the wavelet-based analysis.  Traditional box-counting methods fail both in projection but in the ambient $N$-dimensional space for disconnected sets.  Fourier methods are also ineffective as the Fourier power spectrum is dominated by the large-scale structure of the remnant.  Here, we demonstrate the technique on simulated data and on  observations of Cas A, where we find that  dimension of the measure \textcolor{black}{depends on the frequency observed.  The observations at different frequencies are tracing gas at different temperatures.  }

We believe this approach is quite general and can be used  not only to extract additional information about the underlying physics for a wide range of astrophysical objects, but more generally to analyze data in projection across a variety of fields. \textcolor{black}{ For example, it would be interesting to apply these methods to measurements of the intrinsic dimensions of surfaces in high dimensional spaces in machine learning
\cite{Levina2004}  and in measurements of the fractal dimension of decision boundaries. }

\section{Acknowledgments}

We thank Tea Temim for providing the Cas A image, Guy David for helpful insights on measuring level sets with wavelets in the presence of randomness, and Drummond Fielding and Vikramaditya Giri for many discussions.  We thank Adrian Price-Whelan for assistance with Figure 2.

\bibliography{proj}
\end{document}